# Information-theoretic View of Sequence Organization in a Genome

Liaofu Luo[1*], Yang Gao[1&], Jun Lu[2&]

[1] Laboratory of Theoretical Biophysics, Faculty of Physical Science and Technology, Inner Mongolia University, Hohhot 010021, China;

[2] Department of Physics, Inner Mongolia University of Technology, Hohhot 010051, China.

* Corresponding author.   Email address: lolfcm@mail.imu.edu.cn

& Those authors contributed equally to the work.

## Abstract

Sequence organizations are viewed from two points: one is from informational redundancy or informational correlation (IC) and another is from k-mer frequency statistics. Two problems are investigated. The first is how the ICs exceed the fluctuation bound and the order emerges from fluctuation in a genome when the sequence length attains some critical value. We demonstrated that the transition from fluctuation to order takes place at about sequence length 200-300 thousands bases for human and *E coli* genome. It means that the life emerges from a region between macroscopic and microscopic. The second is about the statistical law of the k-mer organization in a genome under the evolutionary pressure and functional selection. We deduced a sum rule Q(k,N) on the k-mer frequency deviations from the randomness in a N-long sequence of genome and deduced the relations of Q(k,N) with k and N. We found that Q(k,N) increases with length N at a constant rate for most genome sequences and demonstrated that when the functional selection of k-mers is accumulated to some critical value the ordering takes place. An important finding is the sum rule correlated with the evolutionary complexity of the genome.

## 1   Introduction to Informational redundancy: Fluctuation analysis

Once Sir Eddington said: "Suppose that we were asked to arrange the following into two categories: distance, mass, electric force, entropy, beauty, melody, I think that there are the strongest grounds for placing entropy alongside beauty and melody.  Entropy is only found when the parts are viewed in association, and it is by viewing or hearing the parts in association that beauty and melody are discerned." Information is the first important and fundamental concept in life sciences.  However, the Shannon information quantity is essentially the entropy in physics. We have used informational redundancies to describe the nucleic acid sequence [Luo & Li, 1991;Luo et al, 1998]. Define information entropy

$$H = -\sum_i p_i \log_2 p_i \quad (1)$$

and the first -order informational redundancy $D_1$

$$D_1 = H_{max} - H = 2 + \sum p_i \log_2 p_i \quad (2)$$

$D_1$ describes the divergence from equiprobability of base composition of a nucleotide sequence; that is, the deviation of the sequence from a random one.  $D_1$ is related to the vocabulary constitution of genetic language.  Since the information of the genetic language is mainly stored in the base correlation apart from base composition we define Markovian entropy $H_M$ through the conditional probability of adjacent bases, $p_{j|i}$,

$$H_M = -\sum_i p_i \sum_j p_{j|i} \log_2 p_{j|i} \tag{3}$$

and define the second-order informational redundancy $D_2$

$$D_2 = H - H_M = 2H + \sum_{ij} p_{ij} \log_2 p_{ij} \tag{4}$$

Here $p_{ij} = p_i p_{j|i}$ is the joint probability of a pair of bases. $D_2$ describes the deviation of base correlation from independent sequence which is related to the grammatical construction of genetic language. To generalize to non-neighboring case we introduce Markovian entropy with lag

$$H_M^1 = -\sum_i p_i \sum_j p_{j|i*} \log_2 p_{j|i*}$$
$$H_M^2 = -\sum_i p_i \sum_j p_{j|i**} \log_2 p_{j|i**} \tag{5}$$

etc. The star * in $p_{j|i*}$ and $p_{j|i**}$ means any base. Likewise, $D_2$ is also generalized to

$$D_{k+2} = H - H_M^k = 2H + \sum_{ij} p_{i(k)j} \log_2 p_{i(k)j}$$
$$(k = 0, 1, 2 \cdots; \quad H_M^0 = H_M) \tag{6}$$

or

$$D_{k+2} = \sum_{ij} p_{i(k)j} \log_2 \frac{p_{i(k)j}}{p_i p_j} \tag{7}$$

$D_{k+2}$ is called informational correlation (IC). Here $p_{i(k)j}$ means the joint probability of a pair of non-neighboring bases with distance $k$. The equation (7) is identical to Kullback's mutual information. In the calculation of actual sequence with limited length $N$, to reduce the edge effect we assume the boundary condition of periodicity. That is, in calculating $p_{i(k)j}$ ( $p_{i(k)j} = \frac{N_{i(k)j}}{N}$, $N_{i(k)j}$ means the number of base pairs $ij$ with distance $k$) we join the first $k+1$ symbols (bases) onto the tail of the sequence to form a sequence with length $N+k+1$. It is easily proved that under this boundary condition we have

$$H \geq H_M^k, \quad D_{k+2} \geq 0 \quad (k=0,1,2...) \tag{8}$$

The equality occurs as $p_{i(k)j} = p_i p_j$. On the other hand, under periodicity boundary condition one has $N_{i(k)j} = N_{j(N-2-k)i}$, so

$$D_{k+2} = D_{N-k} \tag{9}$$

$(D_1, D_2, \cdots, D_N)$ comprises a set of informational parameters describing base composition and base correlation. For a random sequence of infinite length $(N \to \infty)$, $D_1 = D_2 = \cdots = D_N = 0$; and for an independent sequence of infinite length, $D_2 = D_3 = \cdots = D_N = 0$. However, for a random or independent sequence of limited length there is no relation such as $D_2 = \cdots = D_N = 0$ due to fluctuation effect. So, for finite sequence we should differentiate the base correlation from the contribution of fluctuation.

From Pearson theorem [Feller, 1971], if a set of stochastic variables $(x_1, \cdots, x_m)$ obeys $m$-dimensional normal distribution then the quardratic form obeys $\chi^2$—distribution with $m$ degrees of freedom. The $v$ degrees of freedom $\chi^2$—distribution is given by

$$p(\chi^2, v) = \frac{1}{2^{\frac{v}{2}} \Gamma(\frac{v}{2})} (\chi^2)^{\frac{v}{2}-1} \exp(-\frac{\chi^2}{2}), \quad (\chi^2 \geq 0) \tag{10}$$

Its expectation value and deviation are

$$<\chi^2> = v$$

$$\sigma^2(\chi^2) = 2v \tag{11}$$

When $v \to \infty$, $\chi^2$ distribution approaches to a normal distribution. The $\chi^2$ distribution is commonly used in analyzing the measurement results. If a discrete quantity that can take $m$ values is measured $N$ times, the theoretical frequency of the $i$-th value is $E_i = Np_i (i = 1 \cdots m)$ ($p_i$ denoting the probability of the $i$-th value) and its experimental frequency is $n_i$, $n_i$ — a stochastic variable obeys normal distribution, then the Pearson $\chi^2$ quantity

$$\chi^2 = \sum_i^m \frac{(n_i - E_i)^2}{E_i} \tag{12}$$

obeys $m-1$ degree of freedom $\chi^2$ distribution as $N \to \infty$.

The above formalism can be generalized to multi-dimensional stochastic variables. For two dimensional variables suppose the experimental frequency $n_{ij} (i = 1, \cdots, m_1; j = 1, \cdots, m_2)$,

$$\sum_{ij} n_{ij} = N, \quad \sum_i^{m_1} n_{ij} = N_{\cdot j}, \quad \sum_j^{m_2} n_{ij} = N_{i \cdot}$$

Define Pearson $\chi^2$ quantity

$$\chi^2 = N \sum_{ij} \frac{(n_{ij} - \frac{N_{i \cdot} N_{\cdot j}}{N})^2}{N_{i \cdot} N_{\cdot j}} \tag{13}$$

It obeys $(m_1 - 1)(m_2 - 1)$ degree of freedom $\chi^2$ distribution as $N \to \infty$.

Let us discuss first the statistical distribution of $D_1$ for a sequence of length $N$. For random sequence we have $p_i = \dfrac{1}{4}$. The base number $n_i$ is a stochastic variable, obeying polynomial distribution, or as $N$ large enough, approaching a normal distribution. Set

$$x_i = (n_i - Np_i)/Np_i \tag{14}$$

We have

$$D_1 = \log_2 4 + \sum_i \dfrac{n_i}{N} \log_2 \dfrac{n_i}{N}$$
$$= \dfrac{1}{\ln 2} \sum_i p_i (\dfrac{1}{2} x_i^2 - \dfrac{1}{6} x_i^3 + \dfrac{1}{12} x_i^4 \cdots) \tag{15}$$

So

$$2N \ln 2 D_1 = \sum_i (n_i - Np_i)^2 / Np_i \tag{16}$$

for large $N$. It obeys 3 degrees of freedom $\chi^2$ distribution. We have

$$<D_1> = \dfrac{3}{2N\ln 2}, \quad \sigma^2(D_1) = \dfrac{3}{2N^2(\ln 2)^2} \tag{17}$$

for random sequence as $N \to \infty$. Since $n_i$ obeys polynomial distribution one can estimate the ratio of neglected terms to retained terms in (15) is $\dfrac{5}{12N}$.

Now we discuss $D_2$. For random sequence (or independent sequence), $p_{ij} = p_i p_j$. The base pair number $n_{ij}$ is a stochastic variable. Set

$$x_{ij} = (n_{ij} - Np_i p_j)/Np_i p_j \tag{18}$$

Under periodicity boundary condition we have

$$D_2 = -2 \sum p_i \log_2 p_i + \sum \dfrac{n_{ij}}{N} \log_2 \dfrac{n_{ij}}{N}$$
$$= \dfrac{1}{2\ln 2} \sum_{ij} p_i p_j (x_{ij}^2 - \dfrac{1}{3} x_{ij}^3 + \dfrac{1}{6} x_{ij}^4 + \cdots) \tag{19}$$

For large $N$ case it leads to

$$2N \ln 2 D_2 = \sum_{ij} (n_{ij} - Np_i p_j)^2 / Np_i p_j \tag{20}$$

It obeys 9 degrees of freedom $\chi^2$ distribution. So, for random (independent) sequence we have

$$<D_2> = \frac{9}{2N\ln 2}, \quad \sigma^2(D_2) = \frac{9}{2N^2(\ln 2)^2} \tag{21}$$

as $N \to \infty$. The ratio of neglected terms to retained terms in (19) is $\frac{85}{18N}$.

The statistical distribution of $D_3, D_4, \cdots$ etc in independent sequence is same as $D_2$, obeying 9 degrees of freedom $\chi^2$ distribution, with the expectation and deviation described by (21).

We have shown that the information redundancies $D_1$, $D_2$ etc for random sequence obey $\chi^2$ distribution $p(\chi^2, \nu)$ (Eq (10)). Define

$$\xi = \int_0^{\chi_\xi^2} p(\chi^2, \nu) d\chi^2 \tag{22}$$

as confidence level (C.L.). The upper bound on $\chi^2$ under given confidence level $\xi$ is denoted as $\chi_\xi^2$, which can be found from $\chi^2$ table.

Following previous discussions and $\chi^2$ table we obtain the fluctuation bound (f.b.) of $D_1$ and $D_2$ etc for any real DNA sequence,

$$\begin{aligned} D_1(f.b.) &= \frac{11.3}{2N\ln 2} = \frac{8.15}{N} \quad (99\% \text{ C.L.}) \\ D_{k>1}(f.b.) &= \frac{21.7}{2N\ln 2} = \frac{15.65}{N} \quad (99\% \text{ C.L.}) \end{aligned} \tag{23}$$

This means that for a DNA sequence with $D_1 \geq \frac{8.15}{N}$, the base composition is not at random at 99% confidence level. For a DNA sequence with $D_2 \geq \frac{15.65}{N}$, the neighboring bases occur not independently and the correlation does exist at 99% confidence level. For a DNA sequence with $D_3 \geq \frac{15.65}{N}$, the next-to-neighboring base correlation does exist at 99% confidence level, etc. Oppositely, for DNA sequence with $D_1 < \frac{8.15}{N}$, $D_{k>1} < \frac{15.65}{N}$, these non-vanishing informational redundancies have no biological meaning since their occurrence is due to stochastic fluctuation. Based on above discussions we can split off the fluctuation effect due to finite length.

The informational correlation $D_2$, $D_3$ etc can be rewritten as

$$D_{k+2} \cong \frac{1}{\ln 2} \sum_{ij} \frac{(p_{i(k)j} - p_i p_j)^2}{p_i p_j} \tag{24}$$

The summation in (24) represents 16 kinds of correlation. To describe a particular correlation, define partial informational correlation (PIC)

$$F_{i(k)j} = (p_{i(k)j} - p_i p_j)^2 \tag{25}$$

Introduce

$$G_{i(k)j} = F_{i(k)j} N / p_i p_j (1-p_i)(1-p_j) \tag{26}$$

It can be proved that for random sequence, $G_{i(k)j}$ obeys 1 degree of freedom $\chi^2$ distribution, since

$$G_{i(k)j} = \frac{(n_{i(k)j} - Np_i p_j)^2}{Np_i p_j} + \frac{[n_{\bar{i}(k)j} - N(1-p_i)p_j]^2}{N(1-p_i)p_j}$$
$$+ \frac{[n_{i(k)\bar{j}} - Np_i(1-p_j)]^2}{Np_i(1-p_j)} + \frac{[n_{\bar{i}(k)\bar{j}} - N(1-p_i)(1-p_j)]^2}{N(1-p_i)(1-p_j)} \tag{27}$$

where

$$n_{\bar{i}(k)j} = n_j - n_{i(k)j}, \quad n_{i(k)\bar{j}} = n_i - n_{i(k)j},$$

$$n_{\bar{i}(k)\bar{j}} = N - (n_i - n_{i(k)j}) - (n_j - n_{i(k)j}) - n_{i(k)j}$$

Eq (27) takes the standard form of Pearson quantity, Eq (13), with $m_1 = m_2 = 2$, so it obeys 1 degree of freedom $\chi^2$ distribution. Therefore, the fluctuation bound of $F_{i(k)j}$ is

$$F_{i(k)j}(f.b.) = \frac{6.63}{N} p_i p_j (1-p_i)(1-p_j) \quad (99\% \text{ C.L.}) \tag{28}$$

**Note 1**  Recently Bauer, Schuster and Sayood (2008) proved that the Average Mutual Information (AMI) profiles are species specific and can be looked as a genomic signature. In fact, AMI is informational correlation $D_{k+2}$ defined above and widely used by us in literatures [Luo & Li, 1991; Luo et al, 1998; Luo, 2004]. We have shown that the efficiency of IC to signify a genome is much improved by association with PIC. The point will be published elsewhere [Gao et al, to be published].

**Note 2**  The fluctuation analysis of IC and PIC was published firstly by Luo, Lee, Jia et al (1998). However, this analysis has not been pushed further in bioinformatics application. Here, in this section, we gave a more detailed deduction and explanation of the fluctuation bound of IC and PIC which is taken from Luo's book [Luo, 2004]. The section may serve as an introduction to the problem of informational order of genome.

## 2 How the order of informational correlation of a genome emerges from fluctuation?

For most gene sequences there exists a main maximum located on $k$=0 and 1 neighbors on $D_{k+2} - k$ plot and the nucleotide correlations in the nearest-neighboring and next-to-nearest neighboring sites are dominant ones. The feature is more notable for coding sequences since the correlation with a short tail occurs more frequently on their $D_{k+2} - k$ plot. So, against the complex background of long-range interactions between nucleotides, there exists a definite simplicity – the strong short-range correlation of adjacent bases in a gene. We call the above rule as the short-range dominance of nucleotide correlation in DNA sequence [Luo & Li, 1991; Luo *et al*, 1998]. The short-range dominance of nucleotide correlation affords a sound basis for the application of Markov chain model or hidden Markov chain model to the DNA bioinformatics.

The short-range dominance of nucleotide correlation means the general decreasing trend of informational correlation $D_k$ on $D_{k+2} - k$ plot for a gene sequence of given length. So, apart from some peaks the correlation $D_k$ for most genes lowers with $k$ and takes a value below the fluctuation bound at some large $k$. But, how the picture is retained or changed for a genome where the sequence length is much larger than a gene? Since the fluctuation bound depends on sequence length $N$ as seen from Eq (23), although many $D_k$'s may lose their meaning in a gene scale (i.e. lower than the fluctuation bound of a gene) they always exceed the fluctuation bound and regain the meaning in a genome. This provides a mechanism for the ordering of statistical correlation in a genome as a whole.

Consider how the order emerges when the sequence is growing. Denote $D_k$ of an $N$-long sequence starting from $I$ in a genome as $D_k(N, I)$. The informational correlation has meaning at a definite confidence level when it exceeds the fluctuation bound, namely

$$D_k(N, I) - \frac{b}{N} \geq 0 \quad (k>1)$$
$$b = 15.65 \quad (99\% \text{C.L.})$$
$$\phantom{b =} 20.125 \quad (99.9\% \text{C.L.})$$

(29)

We find for given $k$ and $I$ there always exists a turning point $N_{ck}(I)$ for a genome sequence. That is, Eq (29) always holds for any chromosome sequence starting from $I$ when $N > N_{ck}(I)$ where $N_{ck}(I)$ is the maximal zero point of $N$ satisfying $D_k(N, I) - \frac{b}{N} = 0$. The existence of turning point $N_{ck}(I)$ is due to $D_k(N, I)$ slightly changed with $N$ while the fluctuation bound decreasing with $N$.

Consider sequences with $I$=0 first. For *E coli* (K 12 strain) and human Chr 22 genome sequence. By computing $D_k(N, I = 0)$ and comparing it with the fluctuation bound (99.9% C.L.) we obtain the turning point $N_{ck}(I=0)$ as follows

Table 1  Turning point from fluctuation to order ($N_{ck}$) for *E coli* genome

|  | $D_{50}$ | $D_{200}$ | $D_{350}$ | $D_{500}$ | $D_{650}$ | $D_{800}$ | $D_{1250}$ |
|---|---|---|---|---|---|---|---|
| $N_{ck}$(I=0) (99.9%C.L.) | $4\times10^4$ | $6\times10^4$ | $9\times10^4$ | $9\times10^4$ | $2\times10^5$ | $4\times10^5$ | 4639k |
| $N_{ck}$(I=0) (99% C.L.) | $3\times10^4$ | $4\times10^4$ | $7\times10^4$ | $9\times10^4$ | $2\times10^5$ | $4\times10^5$ | 4639k |

Table 2  Turning point from fluctuation to order ($N_{ck}$) for human genome

|  | $D_{50}$ | $D_{200}$ | $D_{1200}$ | $D_{1400}$ | $D_{3000}$ |
|---|---|---|---|---|---|
| $N_{ck}$(I=0) (99.9%C.L.) | $3\times10^4$ | $4.5\times10^4$ | $5\times10^4$ | $6\times10^4$ | $3\times10^5$ |
| $N_{ck}$(I=0) (99% C.L.) | $2.5\times10^4$ | $3\times10^4$ | $4\times10^4$ | $5\times10^4$ | $2\times10^5$ |

ECO_K12 sequence data taken from http://www.ncbi.nlm.nih.gov/projects/genome/, accession NO. NC_000913; Human chr22, accession NO. NC_000022.

The like analyses have been done for other genomes. We found that the turning point from fluctuation to order does exist for any genome.  From Table 1 and 2 we found that $N_{ck}$ keeps near a constant but slightly increases with *k*.  For human $N_{ck}$ changes with *k* more slowly than *E coli*. This is due to $D_k$ decreasing with *k* more quickly for Hum than for ECO. In general, due to the rapid decrease of $D_k$ one may assume that only informational correlations from *k*=2 to 200 are important. The contribution of neglected higher-order correlations (*k*>200) is in the order of 1/100.

The threshold of *k*, $k_T$ =200, can be approximately assumed as a constant for different genomes. The point can be cleared up by the following observations of $D_k(N,I)$ at given *N*.

Consider the difference between $D_k$'s.  The threshold of *k* is supposed to be determined by the following condition: if

$$|D_k(N,I) - D_{k+3n}(N,I)| < \frac{b}{N} \tag{30}$$

(3*n* introduced to account for reading frame effect, *n*=1，2，3…..) for all $k > k_T$, then $k_T$ is the threshold of {$D_k$}.  Taking $N=10^5$, *I* stochastic and *b*=20.125, we obtain $k_T$ for different species in Table 3. It shows that $k_T \sim 200$ is basically in consistency with all species.

To find a full solution of the problem of the transition from fluctuation to order we should study the dependence of $D_k(N,I)$ on *I*.  For given *k* smaller than 200 and given *N* smaller than $3\times10^5$ in a genome sequence we calculate $D_k(N,I)$ 100 times by randomly sampling *I* and record the number of sequences with $D_k$ higher than fluctuation bound (99.9% CL). The fraction of sequences with $D_k$> f.b. (99.9% CL) is called order fraction (OF). The results are given in Table 4 and 5 for *E coli* and human genome respectively which show how the emergence of order depends on sequence length. In Fig 1 and 2 the order fraction (OF) *vs N* is plotted for each $D_k$, k=50，101，150，200. Each curve is growing with *N* and then saturated．For *E coli*, the fluctuation-to-order transition occurs from $N = 5\times 10^4$ (OF~50%) to $N = 2\times 10^5$ (OF near 100%). For human, the fluctuation-to-order transition occurs from $N = 4\times 10^4$ (OF~50%) to $N = 3\times 10^5$ (OF=99%). From these calculations we conclude that the fluctuation-to-order transition takes place generally at $N = (4\sim 5)\times 10^4$ to $(2\sim 3)\times 10^5$ for a genome.

Table 3  The threshold of $k$ in different genomes

| Species | $K_T$(99.9%) | Species | $K_T$(99.9%) |
|---|---|---|---|
| *M. genitalium* | 269 | *D. melanogaster* | 241 |
| *A. fulgidus* | 260 | *G. gallus* | 73 |
| *E. coli* | 242 | *D. rerio* | 193 |
| *S. cerevisiae* | 202 | *M. musculus* | 187 |
| *A. thaliana* | 199 | *H. sapiens* | 181 |
| *C. elegans* | 157 | | |

The sequence data of *H. sapiens*, *M. musculus* and *G. gallus* are taken from chromosome 1, *D. rerio* from chromosome 5 and others from the whole genome, http://www.ncbi.nlm.nih.gov/projects/genome/.

Table 4  Dependence of order emergence on sequence length for *E coli* genome

| Sequence length | Number of $D_{50}$> f.b. in 100 I's | Number of $D_{101}$> f.b. in 100 I's | Number of $D_{150}$> f.b. in 100 I's | Number of $D_{200}$> f.b. in 100 I's |
|---|---|---|---|---|
| $4\times 10^3$ | 0 | 0 | 2 | 2 |
| $6\times 10^3$ | 2 | 6 | 6 | 4 |
| $8\times 10^3$ | 8 | 8 | 6 | 8 |
| $1\times 10^4$ | 6 | 4 | 2 | 10 |
| $2\times 10^4$ | 26 | 10 | 12 | 4 |
| $3\times 10^4$ | 40 | 22 | 26 | 16 |
| $4\times 10^4$ | 60 | 48 | 36 | 28 |
| $5\times 10^4$ | 58 | 52 | 38 | 36 |
| $6\times 10^4$ | 76 | 62 | 52 | 38 |
| $7\times 10^4$ | 90 | 70 | 74 | 56 |
| $8\times 10^4$ | 96 | 84 | 78 | 64 |
| $9\times 10^4$ | 96 | 88 | 84 | 74 |
| $1\times 10^5$ | 94 | 92 | 90 | 76 |
| $1.5\times 10^5$ | 100 | 94 | 98 | 96 |
| $2\times 10^5$ | 100 | 100 | 100 | 98 |
| $2.5\times 10^5$ | 100 | 100 | 100 | 100 |
| $3\times 10^5$ | 100 | 100 | 100 | 100 |

The dependence of order emergence on sequence length is calculated by randomly sampling *I* . *E coli*_K12 sequence data are taken from http://www.ncbi.nlm.nih.gov/projects/genome/, accession NO. NC_00091.

Table 5  Dependence of order emergence on sequence length for human genome

| Sequence length | Number of $D_{50}>$ f.b. in 100 $I$'s | Number of $D_{101}>$ f.b. in 100 $I$'s | Number of $D_{150}>$ f.b. in 100 $I$'s | Number of $D_{200}>$ f.b. in 100 $I$'s |
|---|---|---|---|---|
| $4 \times 10^3$ | 18 | 12 | 13 | 16 |
| $6 \times 10^3$ | 29 | 16 | 17 | 20 |
| $8 \times 10^3$ | 40 | 19 | 18 | 20 |
| $1 \times 10^4$ | 49 | 21 | 20 | 21 |
| $2 \times 10^4$ | 80 | 33 | 26 | 27 |
| $3 \times 10^4$ | 93 | 48 | 35 | 36 |
| $4 \times 10^4$ | 98 | 59 | 42 | 45 |
| $5 \times 10^4$ | 99 | 69 | 50 | 50 |
| $6 \times 10^4$ | 100 | 77 | 58 | 57 |
| $7 \times 10^4$ | 100 | 81 | 64 | 62 |
| $8 \times 10^4$ | 100 | 86 | 68 | 67 |
| $9 \times 10^4$ | 100 | 88 | 73 | 73 |
| $1 \times 10^5$ | 100 | 92 | 78 | 78 |
| $1.5 \times 10^5$ | 100 | 98 | 89 | 91 |
| $2 \times 10^5$ | 100 | 100 | 96 | 95 |
| $2.5 \times 10^5$ | 100 | 100 | 98 | 97 |
| $3 \times 10^5$ | 100 | 100 | 99 | 99 |

The dependence of order emergence on sequence length is calculated by randomly sampling $I$. Human chr22 sequence data are taken from http://www.ncbi.nlm.nih.gov/projects/genome/, accession NO. NC_000022.

Figure 1  Order fraction (OF) *vs* sequence length (*N*) for *E coli* genome

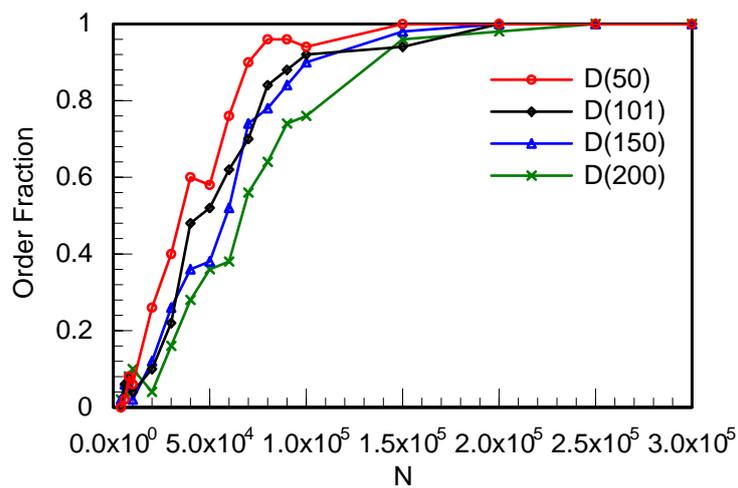

Figure 2   Order fraction (OF) *vs* sequence length (*N*) for human genome

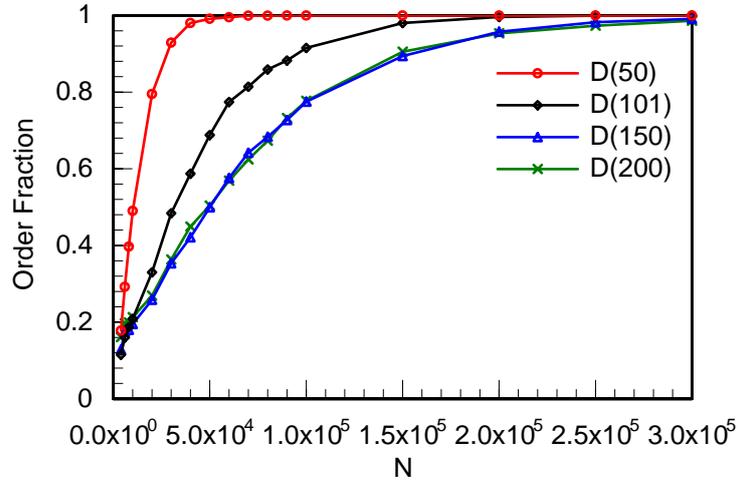

## 3   A non-random k-mer frequency sum rule relating to evolution

The DNA sequence analysis can be carried out through k-mer statistics. Apart from informational redundancies the k-mer frequency distribution is also characteristic of a genome. Recently, the k-mer frequency as a genome barcode was proposed in literatures [Zhou et al, 2008].   However, the set of k-mer frequency parameters is too large for characterizing a genome. In fact, as k=15, the number of k-mers $4^k$ has attained the length of human chromosome. Several years ago we have pointed out that the evolutionary tree can be deduced by non-alignment method – k-mer frequency statistics. By using 16S rRNA (18S rRNA) as molecular clock, we found that the phylogenetic trees deduced from k-mer frequency agree well with life tree when k≥7 [Luo, 2004]. The choice of *k* from 7 to 15 is consistent with a recent analysis by Sims et al [2009]. The k-mer distance method (the conservation of the distance between a pair of k-mers) proposed in human promoter prediction is another new development of the application of k-mer frequency analysis in bioinformatics [Lu et al, 2008].

The evolutionary variation of a genome includes many factors: single base mutation, deletion, insertion, block mutation, recombination, sequence duplication, etc. Accompanying genome variation the distribution of k-mer frequency also changes in the sequence. If the genome growth is a fully stochastic process then DNA should be a random sequence and the k-mer frequency in the sequence takes Poisson distribution with the most probable value

$$\alpha(k) = \frac{N}{4^k} \tag{31}$$

where *N* is genome length.   However, the non-random force does exist in genome evolution. The most important non-random forces are functional selection and sequence duplication. Due to functional selection and sequence duplication the k-mer frequency deviates from Poisson distribution . Set $f_i^k$ is the frequency of the *i*-th k-mer. The square deviation $\sigma_\alpha^2$ is expressed by

$$\sigma_\alpha^2(k) = \sum_i \frac{(f_i^k - \alpha)^2}{4^k} \tag{32}$$

If the peak of k-mer frequency distribution in a genome has shifted to $\beta$ one can easily prove that the frequency deviation around $\beta$ is

$$\sigma_\beta^2 = \sigma_\alpha^2 + (\alpha - \beta)^2 \tag{33}$$

If there exist several peaks in k-mer frequency distribution for a genome (for example, in *G. gallus*) we proved that Eq (33) holds for each peak. So $\sigma_\alpha^2$ is a well-defined quantity which has fundamental meaning for genome analysis.

The ratio of the deviation $\sigma_\alpha^2(k, N)$ for a genome sequence of length $N$ to the deviation $\alpha(k, N)$ for a random sequence of same length represents the nucleotide correlation strength or non-randomness level of the genome which describes how the k-mer frequency distribution of the genome sequence is deviated from the stochastic value due to functional selection and sequence duplication. We denote the ratio as non-randomness parameter $Q(k, N)$

$$Q(k, N) = \frac{\sigma_\alpha^2(k, N)}{\alpha(k, N)} \tag{34}$$

Formally, $Q(k, N)$ is a sum rule over k-mer frequencies. We call the quantity as non-random k-mer frequency sum rule, which reflects the non-randomness of the genome.

Two factors lead to $Q(k, N)$ larger than unity. For a long-$N$ random sequence, as said above, the k-mer frequencies obey Poisson distribution with mean and deviation both equal $\frac{N}{4^k} = \alpha$. However, for a real genome the deviation of k-mer frequency $\sigma_\alpha^2$ is much larger than a random sequence with same length. This is due to functional selection making some k-mer's occurrence much higher than $\alpha$ and some much lower than $\alpha$ in the genome. The stronger the selection, the larger the deviation $\sigma_\alpha^2$.

The another factor is sequence duplication. For a genome of length $N$, to obtain the k-mer frequency deviation same as a randomized genome sequence, it is enough to take only a sequence segment of length $M$ with

$$\frac{M}{N} = \frac{\alpha}{\sigma_\alpha^2} \qquad (M \ll N) \tag{35}$$

This indicates that many segment repetitions exist in the genome. So, $\frac{\sigma_\alpha^2}{\alpha}$ is also a measure of segment repetition, reflecting the self-similarity structure of genome sequence. As estimated by Hsieh et al (2003), the single base mutation number is $4M$ for the genome and the ratio of the mutation to segment repetition is

$$\eta = \frac{4M}{F} \tag{36}$$

($F = \frac{N}{\lambda}$) where $F$ is the number of segment repetition, and $\lambda$ the average repetitious segment length. It leads to

$$\lambda = \frac{1}{4} \frac{\sigma_\alpha^2}{\alpha} \eta \tag{37}$$

The above discussions show that both functional selection and sequence duplication make $Q(k, N)$ growing in the genome evolution.

Next, we study the relation of $Q(k, N)$ with $k$. Define

$$R(k, N) = \frac{Q(k, N)}{Q(k+1, N)} \tag{38}$$

Suppose (k+1)-mer forms from a k-mer and a base independently and the additional base occurs in the sequence with probability 1/4. Introduce probability $p_i^{(k)} = \frac{f_i^k}{N}$. One has $p_j^{(k+1)} \approx p_i^{(k)} p_{i'}$, $p_{i'} \approx 1/4$. So,

$$\begin{aligned}\sigma_\alpha^2(k+1) &= \sum_j \frac{(f_j^{k+1} - \alpha(k+1))^2}{4^{k+1}} \\ &\cong N^2 \sum_i^{4^k} \sum_{i'}^{4} (p_i^{(k)} p_{i'} - \frac{1}{4^{k+1}})^2 / 4^{k+1} \\ &\cong \frac{1}{16} \sigma_\alpha^2(k)\end{aligned} \tag{39}$$

It leads to $R(k, N) = 4$ approximately. In fact, the extension of k-mer to (k+1)-mer is not an statistically independent process. Due to correlation existing between the k-mer and the (k+1)-th base in the (k+1)-mer, one has

$$\frac{1}{4} \sum_i^{4^k} (p_i^{(k)} - \frac{1}{4^k})^2 < \sum_j^{4^{k+1}} (p_j^{(k+1)} - \frac{1}{4^{k+1}})^2 < \sum_i^{4^k} (p_i^{(k)} - \frac{1}{4^k})^2 \tag{40}$$

Therefore,

$$1 < R(k, N) < 4 \tag{41}$$

Along with $Q(k, N)$ describing correlation strength, $R(k, N)$ describes the steepness of correlation strength. The calculation results of $Q(k, N)$ and $R(k, N)$ for typical sequences of *E coli* (K12 strain) and human Chr 22 are given in Table 6 and 7 respectively. For *E coli*, the variation of $Q(k,N)$ for different sequences (sequences with different starting base $I$ in genome) is about 10% or smaller. For human, the variation of $Q(k,N)$ is also in the range of 10% for $N>5\times 10^6$; but due to the existence of large amount of simple repetitious segments $Q(k,N)$ can deviate from its typical average value by a factor 2 or more for some sequences of $N<5\times 10^6$, and even deviate by a factor higher than 20 for some sequences of $N<10^6$.

Table 6　$Q(k,N)$ and $R(k,N)$ for E coli genome

| $N$ | $4.64\times 10^6$ | $10^6$ | $5\times 10^5$ | $10^5$ | | $4.64\times 10^6$ | $10^6$ | $5\times 10^5$ | $10^5$ |
|---|---|---|---|---|---|---|---|---|---|
| $Q(3,N)$ | 5390 | 1240 | 622 | 146 | $R(3,N)$ | 2.10 | 2.11 | 2.12 | 2.13 |
| $Q(4,N)$ | 2570 | 589 | 293 | 69 | $R(4,N)$ | 2.51 | 2.51 | 2.51 | 2.46 |
| $Q(5,N)$ | 1020 | 235 | 117 | 28 | $R(5,N)$ | 2.76 | 2.73 | 2.71 | 2.56 |
| $Q(6,N)$ | 371 | 86 | 43 | 11 | $R(6,N)$ | 2.92 | 2.85 | 2.79 | 2.48 |
| $Q(7,N)$ | 127 | 30 | 15 | 4.4 | $R(7,N)$ | 3.01 | 2.84 | 2.66 | 2.1 |
| $Q(8,N)$ | 42 | 11 | 5.8 | 2.1 | $R(8,N)$ | 2.94 | 2.59 | 2.23 | 1.5 |

Table 7　$Q(k,N)$ and $R(k,N)$ for human genome

| $N$ | $7.5\times 10^6$ | $10^6$ | $5\times 10^5$ | $10^5$ | | $7.5\times 10^6$ | $10^6$ | $5\times 10^5$ | $10^5$ |
|---|---|---|---|---|---|---|---|---|---|
| $Q(3,N)$ | 17700 | 3170 | 1560 | 319 | $R(3,N)$ | 2.30 | 2.49 | 2.49 | 2.48 |
| $Q(4,N)$ | 7690 | 1270 | 629 | 129 | $R(4,N)$ | 2.52 | 2.58 | 2.61 | 2.66 |
| $Q(5,N)$ | 3050 | 492 | 241 | 48 | $R(5,N)$ | 2.55 | 2.52 | 2.56 | 2.63 |
| $Q(6,N)$ | 1200 | 196 | 94 | 18 | $R(6,N)$ | 2.36 | 2.26 | 2.27 | 2.43 |
| $Q(7,N)$ | 508 | 86 | 42 | 7.6 | $R(7,N)$ | 2.06 | 1.91 | 1.90 | 2.07 |
| $Q(8,N)$ | 247 | 45 | 22 | 3.7 | $R(8,N)$ | 1.73 | 1.58 | 1.58 | 1.64 |

From Table 6 and 7 we find the correlation strengths $Q(k,N)$ for human are higher than those for E coli but there is little difference for the steepness $R(k,N)$ between two species. For given genome and given $N$ the steepness $R(k,N)$ increases with $k$ then lowers down through a maximum at $k=4$ to 6. $\frac{dR_k}{dk}=0$ occurring at $k = 4$ to 6 is related to the fact that a large portion of functional segments in the genome are 4-mers, 5-mers and 6-mers and these k-mer frequencies are sensitively dependent of $k$.

The correlation strength $Q(k,N)$ depends on sequence length. As seen from two tables $Q(k,N)$ is an increasing function of $N$. The point can be proved by the following calculation.

$$\frac{dQ(k,N)}{dN} = \frac{d}{dN}\left(\frac{\sum_i (f_i^k)^2}{N}\right) - \frac{1}{4^k}$$
$$= \sum_i (p_i^{(k)})^2 - \frac{1}{4^k} + N\frac{d\sum_i (p_i^{(k)})^2}{dN} \tag{42}$$

where $f_i^k(N) = p_i^{(k)} N$ has been introduced. For most genome sequences,

$$N\frac{d\sum_i (p_i^{(k)})^2}{dN} \quad \text{and} \quad N\frac{d^2\sum_i (p_i^{(k)})^2}{dN^2}$$

are small quantity when $N$ large. So, by use of

$$\frac{1}{4^k} < \sum_i (p_i^{(k)})^2 < 1$$

we have

$$0 < \frac{dQ(k,N)}{dN} < (1 - 4^{-k}) \tag{43}$$

and

$$\frac{d^2 Q(k.N)}{d^2 N} = 0 \tag{44}$$

for most genome sequences when $N$ large. Eqs (43) and (44) are two noticeable results. Eq (43) indicates that the correlation strength $Q(k,N)$ increases with $N$ and Eq (44) indicates that the increase of $Q(k,N)$ remains a constant rate in a genome.

The changes of $Q(k,N)$ with $N$ in the range of $N=4\times 10^3$ to $2\times 10^5$ for *E coli* and human are plotted in Figure 3 and 4 respectively. The figures are plotted for typical sequences without simple repetitious segments. $Q(k,N)$ is larger than $Q(k+1,N)$ which is in accordance with $R(k,N) \sim 2$ to 3 as seen from Table 6 and 7. The slope of $Q(4,N)$ with $N$ is about $1\times 10^3$ for human and $0.5\times 10^3$ for *E coli*. Figures 3 and 4 show clearly that $Q(k,N)$ increases with $N$ in the range of sequence where the fluctuation-to-order transition occurs. That is, the non-random k-mer frequency sum rule $Q(k,N)$ grows to some critical value so that the ordering takes place.

Figure 3  Change of $Q(k, N)$ with $N$ for typical *E coli* sequence

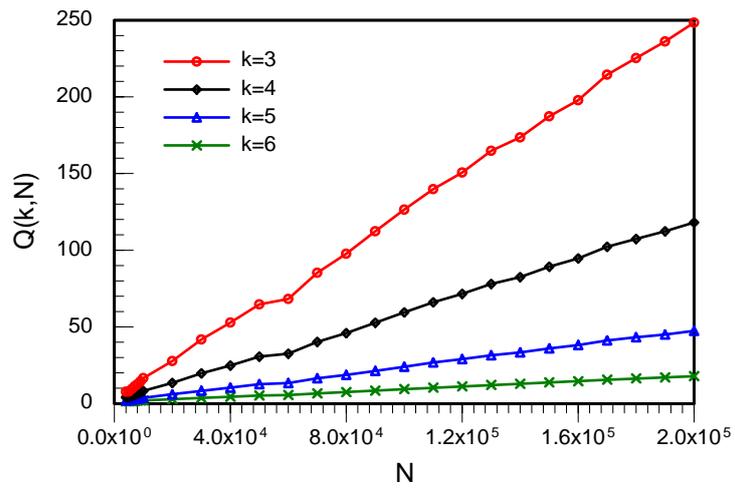

Figure 4  Change of $Q(k, N)$ with $N$ for typical human sequence

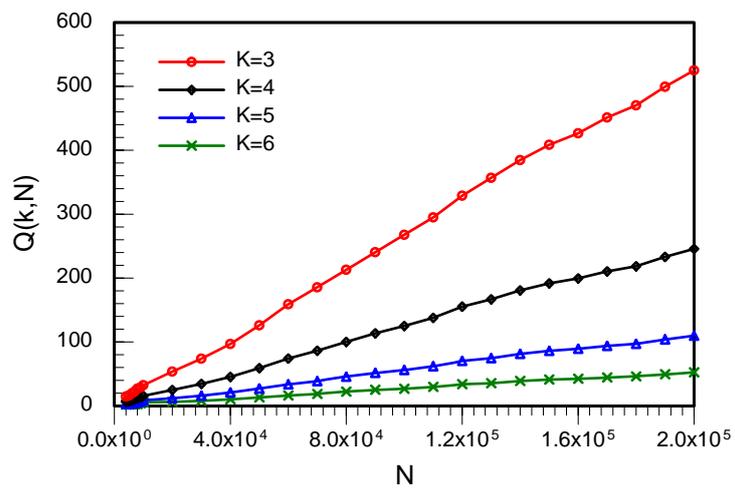

The dependence of correlation strength $Q(k,N)$ on species has been calculated for $k=6$ and $N=N_f$ where $N_f$ is the length of genome, i.e. the total length of all chromosomes. The results are summarized in Table 8. From Table 8 we find there exists a meaningful relation between correlation strength and genome complexity. The relation occurs partly due to the sequence length growing in evolution. For a multi-chromosome genome we found $Q(6,N_c)$ ($N_c$—chromosome length) of different chromosomes is approximately proportional to $N_c$. For example, for 24 chromosomes of human genome, the ratio of $Q(6,N_c)$ to $N_c$ remains a constant, $(1.6 \sim 1.8) \times 10^{-4}$ per base. So, $Q(6,N_f)$ for a genome can be calculated by the sum of $Q(6,N_c)$ for all chromosomes.

Table 8  The relation between $Q(6,N_f)$ and species

| Species | $N_f$ | $Q(6,N_f)$ |
| --- | --- | --- |
| M. genitalium | 580076 | 244 |
| A. fulgidus | 2178400 | 212 |
| E. coli | 4639677 | 371 |
| S. cerevisiae | 12070900 | 1642 |
| A. thaliana | 118960610 | 23479 |
| C. elegans | 100269917 | 40371 |
| D. melanogaster | 72736145 | 7854 |
| G. gallus | 285755532 | 72776 |
| D. rerio | 859083401 | 133051 |
| M. musculus | 2617133082 | 389371 |
| H. sapiens. | 2893618752 | 494841 |

(Sequence data taken from http://www.ncbi.nlm.nih.gov/projects/genome/)

## 4  Concluding remarks

In his book "A Physicist Looks at the Life" (in Chinese, 1994) Luo proposed that the life emerges from a region between macroscopic and microscopic. " How did the biological order emerge from a physical world, or how was life information accumulated for a primitive life?  This question still remains a riddle at present time. Theoretical calculation shows that the mitosis cannot progress successfully if its radius is smaller than $4 \times 10^{-6}$ cm.  A cell with comparatively perfect function should at least have a diameter of $10^{-5}$ cm.  On the other hand, a cell must exchange material and energy with environment or adjacent cells through its membrane. The metabolism of a cell cannot progress if it has not enough surface area.  But the ratio of surface area to volume is inversely proportional to the diameter.  A small enough diameter is necessary for a cell to maintain the equilibrium between nucleus and cytoplasm.  So, the smallest basic unit of life should have a size of $10^{-5}$ cm or larger. This size falls between macroscopic and microscopic dimensions. On the other hand, if a macroscopic dimension is marked by Avgadero constant $N=6 \times 10^{23}$ and a microscopic dimension is marked by 1, then the dimension of a genome is nearly $\sqrt{N}$, which falls between macroscopic and microscopic exactly." The above discussions in section 2 show that the informational correlation becomes ordered only when the genome size exceeds 200~300 thousands bases. It affords a further evidence on the life emerging from a region between macroscopic and microscopic.

$D_k(N, I)$ is an appropriate quantity to discuss the fluctuation – order transition of a genome. The present work reveals the existence of fluctuation – order transition in a genome and explores its statistical characters. However, it has not been given the mechanism of the transition. Genome is the fundamentals of life where the most important interaction is DNA sequence encoding protein and the protein molecule recognizing and being bound with DNA sequence. So, we should use the di-sequence interaction model of DNA and protein instead of the single DNA sequence model. This is what we shall do. It is expected that after introducing the interaction between protein and DNA, the fluctuation – order transition will show more characteristics of "phase transition" at the transition point as those occurring in other typical phase transitions in physics.

Apart from information redundancies the k-mer frequencies give another approach to studying the genome sequence organization from the point of bioinformatics. To avoid the difficulty in using too many parameters in this approach, in section 3 we proposed a sum rule of k-mer frequencies. $Q(k,N)$ describes the non-random distribution of $4^k$ kinds of k-mers. It is a good quantity for genome description due to its clear relation with $k$ ($R(k,N)$ ~ 2 to 3) and $N$ ($\frac{dQ(k,N)}{dN}$ ~ constant). The sum rule gives a measure of functional selection and sequence duplication. The increase of $Q(k,N)$ with $N$ at a constant rate shows the homogenization of the ordering of k-mer frequency deviation in a genome. Due to $Q(k,N)$ linearly growing with $N$, the non-random k-mer frequency sum rule grows in the range of sequence where the fluctuation-to-order transition takes place. That is, when the functional selection of k-mer is accumulated to some critical value the ordering takes place. As shown in Figure 3 and 4 the critical value of $Q(4,N)$ is about 100 for *E coli* and 200 for human.

Interestingly, the measure $Q(k,N)$ correlates with the evolutionary complexity of species. The relation can be explained by the functional selections or constraints applied on k-mers strengthened in the course of evolution. From above studies we find that the most efficient way to raise the $Q(k,N)$ level of a genome is to enlarge the genome size and to increase its slope. Therefore, the sequence duplication and the functionalization of repetitious sequence is the most efficient driver for the rapid evolution of genome. Generally speaking, the fate the sequence duplication would have is one of the following: the repetitious sequence has nothing to do with the function and sometimes is eliminated by the evolution, or the repetitious sequence is functionalized and contributes to the coding information quantity of the genome. As Luo indicated, the growing of coding information quantity of genome determines the evolutionary direction. Since both two factors, functional selection and sequence duplication, contribute to the non-randomness of k-mers. The separation of these two factors in $Q(k,N)$ will be helpful to the solution of the directionality of genome evolution [Luo, 2008].